\documentclass[a4paper, 12pt,psfig]{iopart}
\usepackage{iopams}
\usepackage{epsf}
\usepackage{euscript}
\usepackage{amsthm}
\usepackage{setstack}
\usepackage{graphicx}
\usepackage{color}
\usepackage{bm}
\usepackage{savesym}

\makeatletter

\newcommand{\Rmnum}[1]{\expandafter\@slowromancap\romannumeral #1@}
\makeatother

\begin{document}
\title{Periodic Walks on Large Regular Graphs and Random Matrix Theory}
\author{Idan Oren$^{1}$ and Uzy Smilansky$^{1,2}$}
\today
\address{$^{1}$Department of Physics of Complex Systems,
Weizmann Institute of Science, Rehovot 76100, Israel.}
\address{$^{2}$School of Mathematics, Cardiff University, Cardiff,
Wales, UK}
 \ead{\mailto{idan.oren@weizmann.ac.il}\\
 \mailto{\ \ \ \ \ \ \ uzy.smilansky@weizmann.ac.il}}

\begin{abstract}
We study the distribution of the number of (non-backtracking)
periodic walks on large regular graphs. We propose a formula for the
ratio between the variance of the number of $t$-periodic walks and
its mean, when the cardinality of the vertex set $V$ and the period
$t$ approach $\infty$ with $t/V\rightarrow \tau$ for any $\tau$.
This formula is based on the conjecture that the spectral statistics
of the adjacency eigenvalues is given by Random Matrix Theory (RMT).
We provide numerical and theoretical evidence for the validity of
this conjecture. The key tool used in this study is a trace formula
which expresses the spectral density of $d$-regular graphs, in terms
of periodic walks. \\

\end{abstract}

\section{Introduction}
  \label{sec:intro}
A graph $\mathcal{G}$ is a set $\mathcal{V}$ of vertices connected
by a set $\mathcal{E}$ of edges. The number of vertices is denoted
by $V = |\mathcal{V}|$ and the number of edges is $E=|\mathcal{E}|$.
In the present work we deal with connected, \emph{simple graphs}
where parallel edges or loops are not allowed. The $V\times V$
\emph{adjacency} (\emph{connectivity}) matrix $A$ is defined such
that $A_{i,j}=1$ if the vertices $i,j$ are connected and $0$
otherwise. A $d$-regular graph is a simple graph in which every
vertex is connected to exactly $d$ vertices.\\
Let $\mathcal{G}_{V,d}$ be the ensemble of $d$-regular simple graphs
on $V$ vertices. Averaging over this ensemble will be carried out
with uniform probability and will be denoted by $\mathbb{E}(\cdot)$.
\footnote[1]{We exclude the set of non-connected graphs since the
probability of a non-connected graph is exponentially small in
$\mathcal{G}_{V,d}$.}\\
Let $P_t$ be the number of $t$-periodic walks without back-track. It
is known that $\mathbb{E}(P_t)=(d-1)^t$ for $t<\log_{d-1}V$, and
that the numbers of cycles $C_t=\frac{P_t}{2t}$ are distributed as
independent Poisson variables \cite{Bollobas,Janson,Wormald}.
Therefore,
\begin{equation}
\label{eq:firsorder}
\frac{1}{V}\cdot\frac{var(P_t)}{\mathbb{E}(P_t)}=2\tau\ ,\ {\rm
where} \ \ \tau\equiv\frac{t}{V} \rightarrow 0.
\end{equation}
We conjecture the following relation valid for any $\tau$ in the
limit $t,V\rightarrow \infty , \ \ t/V\rightarrow \tau$\ :
\begin{equation}
\label{eq:conjecture0}
\frac{1}{V}\cdot\frac{var(P_t)}{\mathbb{E}(P_t)} = F_{COE}\left(
\tau \right)
\end{equation}
where  $F_{COE}\left( \tau \right)$ is a function derived from
Random Matrix Theory for the Circular Orthogonal ensemble (COE), and
is given explicitly in (\ref{eq:var_to_mean}). It takes the
following asymptotic values:
\begin{eqnarray} \hspace{-10mm}
  \label{eq:conjecture1}
 F_{COE}\left(
\tau \right)= \left\{\begin{array}{lcr}2\tau\left
(1+C(d)\tau^{\frac{1}{2}}+ \mathcal{O}(\tau)\right)
 & \mbox{for} & \tau\rightarrow 0\\
\\ 2  & \mbox{for} & \tau\rightarrow \infty\end{array}\right. \ .\\
C(d) = \frac{(d-2)}{\sqrt{2d(d-1)}} \left(
\frac{2}{\pi}\cdot\text{arccoth}(\sqrt{2})-\frac{2\sqrt{2}}{3\pi}-1
\right).  \nonumber
 \end{eqnarray}

This conjecture stems from our previous work
\cite{IdanAmitUzy,IdanUzy2}. Here we shall briefly review the
essence of the previous results, generalize them and present the
numerical data which substantiate our claim for the validity of
(\ref {eq:conjecture0}) and (\ref{eq:conjecture1}).

To set the scene, we shall define the necessary objects and review
some properties of $d$-regular simple graphs.

\subsection{Definitions}

To any edge $b=(i,j)$ one can assign an arbitrary direction,
resulting in two \emph{directed edges}, $e=(i,j)$ and $\hat
e=(j,i)$. Thus, the graph can be viewed as $V$ vertices connected by
edges $b=1,\cdots,E$ or by $2E$ directed edges $e=1,\cdots,2E$ (The
notation $b$ for edges and $e$ for directed edges will be kept
throughout). It is convenient to associate with each directed edge
$e =(j,i)$ its \emph{origin} $o(e) =i$ and \emph{terminus} $t(e)=j$
so that $e$ points from the vertex $i$ to the vertex $j$. The edge
$e'$ follows $e$ if $t(e)=o(e')$.

A \emph{walk} of length $t$ from the vertex $x$ to the vertex $y$ on
the graph is a sequence of successively connected vertices
$x=v_1,v_2,\cdots,v_t=y$. Alternatively, it is a sequence of $t-1$
directed edges $e_1,\cdots , e_{t-1}$ with $o(e_i)=v_i,\
t(e_i)=v_{i+1}, o(e_1)=x,\ t(e_{t-1})=y$. A \emph{t-periodic walk}
is a walk of $t$ steps which starts and ends at the same vertex. A
walk where $e_{i+1} \ne \hat{e}_i$ will be called a \emph{walk with
no back-track} or an \emph{nb-walk} for short.

In order to count $t$-periodic nb-walks it is convenient to
introduce the $2E\times2E$ Hashimoto matrix $Y$ \cite{hashimoto89}
which describes the connectivity of the graph in terms of its
directed edges, and avoids back-tracking:
\begin{equation}
Y_{e,e'}=\delta_{t(e),o(e')}-\delta_{\hat{e},e'}\ .
\label{eq:Ymatrix}
\end{equation}
The number of $t$-periodic nb-walks is $P_t=\tr Y^t$.

The spectrum $\sigma(A)=\{\mu_j\}_{j=1}^{V}$ consists of the
eigenvalues of $A$. The largest eigenvalue is $\mu_V=d$ and it is
simple for connected graphs. The \emph{spectral measure} (spectral
density), from which we exclude the trivial eigenvalue $\mu_V=d$, is
defined as
\begin{equation}
\rho (\mu)  \equiv \frac{1}{V-1}\sum_{j=1}^{V-1}\delta(\mu-\mu_j)\ .
 \label{eq:density}
\end{equation}
The spectrum of $A$ is divided in two complementary sets. The first,
denoted by $R$, consists  of all the eigenvalues which satisfy:
$\left| \mu_k \right| \le 2\sqrt{d-1}$. The complement, $R^c$
consists of the eigenvalues for which $\left| \mu_k \right| >
2\sqrt{d-1}$. The graph is Ramanujan, if $R^c=\emptyset$.\\
In what follows we shall be interested in the large $V$ limit, and
in most cases the replacement of $V-1$ by $V$ will be justified. We
shall do this consistently to simplify the notation.\\
The trace formula which will be derived in the next section
expresses $\rho(\mu)$ as a sum of two contributions. The first,
often  referred to  as the `smooth part' of the spectral density, is
the celebrated Kesten-McKay measure \cite{Kesten,McKay}:
\begin{equation}
    \label{eq:Mckay}
\rho_{KM}(\mu) = \frac{d}{2\pi}\cdot
\frac{\sqrt{4(d-1)-\mu^2}}{d^2-\mu^2}=\lim_{V \rightarrow
\infty}\mathbb{E}(\rho(\mu)).
\end{equation}
The second part is called the `oscillatory part' (or `fluctuating
part') of the spectral density. It is an infinite sum over periodic
walks on the graph, where each term consists of an amplitude which
is combinatorial in nature, and some phase. Although this part is
small compared to the smooth part, it encodes all the interesting
features of the graphs, including the statistics
of the periodic walks.\\

\section{The Trace Formula}
 \label{sec:tf}
The starting point for the derivation is the Bass Identity
\cite{bass} which, for $d$-regular graph reads:
\begin{equation}
\hspace{-15mm}  \det(I^{(2E)}-sY))=
 (1-s^2)^{E-V}\det(I^{(V)}(1+(d-1)s^2)-sA)\ .
 \label{eq:bartholdi}
 \end{equation}
The parameter $s$ is an arbitrary real or complex number, $I^{(2E)}$
and $I^{(V)}$ are the identity matrices in dimensions $2E$ and $V$,
respectively, and the matrices $A$ and $Y$ were defined above. The
Bass identity implies that the spectrum of the Hashimoto matrix $Y$
(\ref {eq:Ymatrix}) is:
\begin{eqnarray}
   \label{eq:specY}
 \sigma(Y) &=&\left \{(d-1), 1,\ +1\times (E-V),\
-1\times (E-V), \right .\nonumber \\
& & \left . ( \sqrt{d-1}\ {\rm e}^{i\phi_k},\ \sqrt{d-1}\
{\rm e}^{-i\phi_k},\ k =1,\cdots (V-1))\right \} \\
& & {\rm where} \ \ \ \phi_k=\arccos\frac{\mu_k}{2\sqrt{d-1}}\ , \ \
\ \  0\le \ \Re (\phi_k)\ \le \pi. \nonumber \\
 & & \mu_k \in \sigma(A)\setminus \{ d \}. \nonumber
 \end{eqnarray}
We can now write down explicitly $\tr Y^t$ which provides the number
of
t-periodic nb walks  :\\
\begin{equation}
\hspace{-25mm} \tr Y^t = (d-1)^t+2(d-1)^{t/2}\left(\sum_{\mu_k\in
R^c}\cosh\left(t\psi_k \right)+\sum_{\mu_k\in
R}\cos\left(t\phi_k\right)\right)+1+(E-V)\cdot(1+(-1)^t).
\end{equation}
where $\phi_k$ is defined in (\ref{eq:specY}) and
$\psi_k=\text{arccosh}\left(\frac{\mu_k}{2\sqrt{(d-1)}}\right)$.\\
It is convenient to introduce the quantities $y_t$,
\begin{equation}
y_t = \frac{1}{V} \frac{\tr Y^t-(d-1)^t-2(d-1)^{t/2}\sum_{\mu_k\in
R^c}\cosh\left(t\psi_k \right)}{(d-1)^{t/2}}\ .
  \label{eq:y_t}
\end{equation}
Not much is known rigorously about the properties of the
non-Ramanujan component of the spectrum. However, we shall use the
known estimates due to Friedman \cite{Friedman} and Hoory  {\it et.
al.} \cite{Hoory}, to show that $y_t$ are  bounded as $t\rightarrow
\infty$. The largest non-Ramanujan eigenalue is equal to
$2\sqrt{d-1}\cdot(1+\epsilon)$, where $\epsilon$ is proportional to
$V^{-\alpha}$ and $\alpha\approx0.6$. With this estimate, the two
last terms in the nominator of (\ref{eq:y_t}) behave asymptotically
as:
\begin{equation}
\hspace{-25mm} (d-1)^t+2(d-1)^{t/2}\sum_{\mu_k\in
R^c}\cosh\left(t\psi_k
\right)\approx\\
(d-1)^t\cdot\left[ 1+\exp\left( -t\left( \frac{1}{2}\ln(d-1)-C\cdot
V^{-0.3} \right) \right) \right]
\end{equation}
where $C$ is some positive constant. For sufficiently large $V$
$$\frac{1}{2}\ln(d-1)-C\cdot
V^{-0.3}>0$$ and therefore the contribution of the non-Ramanujan
eigenvalues is exponentially small compared to the leading term,
$(d-1)^t$. Thus, since the leading order term of $\tr Y^t$ is
$(d-1)^t$, the $y_t$ are bounded independently of the
graph being Ramanujan or not.\\
The explicit expressions for the eigenvalues of $Y$ are now used to
write,
\begin{eqnarray}
\label{eq:ywt}
 \hspace{-25mm}y_t =
\frac{1}{V}\left(\frac{1}{d-1}\right)^\frac{t}{2}+ \frac{d-2}{2}
\left(\frac{1}{d-1}\right)^\frac{t}{2}
(1+(-1)^t)+\frac{2}{V}\sum_{\mu_k\in R}
T_t(\frac{\mu_k}{2\sqrt{(d-1)}})
\end{eqnarray}
where $T_t(x) \equiv \cos{(t\arccos{x})}$ are the Chebyshev
polynomials of the first kind of order $t$. An algebraic,
straightforward derivation which can be found in \cite{IdanAmitUzy},
results in an expression for  $\rho_R(\mu)$ which is the spectral
density restricted to the interval $|\mu|\le 2\sqrt{(d-1)}$ :
\begin{equation}
\hspace{-25mm} \rho_R(\mu) = \frac{d}{2\pi}\cdot
\frac{\sqrt{4(d-1)-\mu^2}}{d^2-\mu^2}+\frac{1}{\pi}Re\sum_{t=3}^\infty
\frac{y_t}{\sqrt{4(d-1)-\mu^2}} e^{ it \arccos \left(
\frac{\mu}{2\sqrt{(d-1)}}\right)}+\mathcal{O}\left(\frac{1}{V}\right)\
.
   \label{eq:trace formula}
\end{equation}
The first term is the smooth part, and can be identified as the
Kesten-McKay density. We notice that the mean value of $y_t$
vanishes as $\mathcal{O}\left(\frac{1}{V}\right )$. This is because
the counting statistics of $t$-periodic nb-walks with $t <
\log_{d-1} V $ is Poissonian, with $\mathbb{E}\left( \tr Y^t \right)
= (d-1)^t$, and because for larger values of $t$, the leading order
term of $\tr Y^t$  is $(d-1)^t$.\\
As a result:
\begin{equation}
\lim_{V\rightarrow \infty} \mathbb{E}\left( \rho(\mu)\right) =
\rho_{KM}(\mu).
\end{equation}
The above can be considered as an independent proof of the
Kesten-McKay formula (\ref{eq:Mckay}). The original derivation
relied on the fact that $d$-regular graphs look locally like trees,
for which the spectral density is of the form (\ref{eq:Mckay}).
Here, it emerged without directly invoking the tree approximation,
rather, it appeared as a result of an algebraic manipulation. The
second term is the aforementioned oscillatory part, which we shall
take advantage of in the sequel.

The remainder term in the trace formula (\ref {eq:trace formula}) is
known, and an explicit expression is given in \cite{IdanAmitUzy}. We
should mention that (\ref{eq:trace formula}) is identical to a trace
formula derived by P. Mn\"{e}v \cite{Mnev} using a different
approach.

\section{Spectral Fluctuations on Graphs and RMT - Introduction and Numerical
Evidence}

Until recently, the only evidence suggesting a connection between
RMT and the spectral statistics on graphs was the numerical studies
of Jacobson {\it et. al.} \cite{Jacobson}. In a preliminary step in
the present research, we performed numerical simulations which
extended the tests of \cite{Jacobson} (see \cite{IdanUzy2}). While
describing these studies, we shall introduce a few concepts from RMT
which will be used in the main body of the paper.

We shall now (and hereafter), work with the variable $\phi$ rather
than $\mu$, as defined in (\ref{eq:specY}). This change of variables
is well-defined since at this stage we have already taken care of
all the eigenvalues lying outside the Kesten-McKay support,
$[-2\sqrt{d-1},2\sqrt{d-1}]$. For this reason all the numerics in
this paper were carried out using the Ramanujan component of
$\sigma(A)$, and the number of relevant eigenvalues is modified
accordingly by
$V\rightarrow V-r_c$, $r_c$ being the cardinality of $R^c$.\\

The Kesten-McKay density mapped onto the circle is not uniform:
\begin{equation}
\rho_{KM}(\phi) = \frac{2(d-1)}{\pi d}\frac{\sin^2\phi}
{1-\frac{4(d-1)}{d^2}\cos^2\phi}\ .
  \label{eq:Kesten McKay phi}
\end{equation}
The mean spectral counting function provides the average number of
eigenvalues up to a certain value. It is defined as
\begin{equation}
\hspace{-17mm} N_{KM}(\phi)=V \int_0^{\phi }\rho_{KM}(\phi) d\phi\ =
V\frac{d}{2\pi}\left(\phi-\frac{ d-2 }{d}\arctan\left
(\frac{d}{d-2}\tan\phi\right )\right). \label{eq:counting}
\end{equation}

Following the standard methods of spectral statistics, one
introduces a new variable $\theta$, which is uniformly distributed
on the unit circle. This ``unfolding" procedure is explicitly given
by
\begin{equation}
\theta_j= \frac{2\pi}{V}N_{KM}(\phi_j)
  \label{eq: definition theta_j}
\end{equation}

\begin{figure}[h]
  \centering
  \scalebox{0.8}{\includegraphics{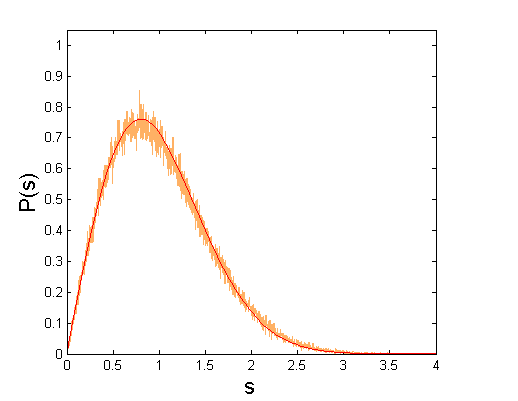}}
 \caption{Nearest level spacings for 3-regular graphs with 1000 vertices.\\
 The figure is accompanied with the RMT prediction for the COE.}
 \label{fig:Nearest Level Spacings}
\end{figure}

The nearest spacing distribution defined as
\begin{equation}
P(s) = \lim_{V\rightarrow\infty}\ \frac{1}{V} \mathbb{E}\left(
\sum_{j=1}^{V}\delta \left
(s-\frac{V}{2\pi}(\theta_j-\theta_{j-1})\right)\right),\
(\theta_0=\theta_V),
\end{equation}
is often used to test the agreement with the predictions of RMT
(this was also the test conducted in \cite{Jacobson}). In figure
(\ref{fig:Nearest Level Spacings}) we show numerical simulations
obtained by averaging over 1000 randomly generated $3$-regular
graphs on $1000$ vertices together with the predictions of RMT for
the COE \cite {Haakebook}. The agreement is quite impressive.

Another quantity which is often used for the same purpose is the
spectral form-factor. This quantity is the main function which we
make use of in this paper. It is given by
\begin{eqnarray}
K_{V}(t)&=&\frac{1}{V}\mathbb{E}\left( \left|\sum_{j=1}^{V}{\rm
e}^{it\theta_j}\ \right|^2\ \right) = 1+\frac{1}{V}\mathbb{E}\left(
\sum_{i\ne j}^{V}{\rm e}^{it(\theta_i-\theta_j)}\ \ \right) \nonumber\\
& & = 1+\frac{2}{V}\mathbb{E}\left( \sum_{i< j}^{V}\cos
t(\theta_i-\theta_j)\ \ \right) \ .
  \label{eq:form factor definition}
\end{eqnarray}
The form-factor is the Fourier transform of the spectral two point
correlation function. It plays a very important r\^ole in
understanding the relation between RMT and the quantum spectra of
classically chaotic systems \cite{Haakebook, BerryA400}.

In RMT the form factor displays scaling. In the limit
$V,t\rightarrow \infty\ \ ;\ t/V = \tau$:
$$K_V(t)=K(\tau \equiv \frac{t}{V})$$
The explicit limiting expressions for the COE ensemble is \cite
{Haakebook}:
\begin{eqnarray} \hspace{-10mm}
\label{eq:K_COE} K_{COE}(\tau) =
\left\{\begin{array}{lcr}2\tau-\tau\log{(2\tau+1),} & \mbox{for} & \tau<1\\
\\ 2-\tau\log{\frac{2\tau+1}{2\tau-1},} & \mbox{for} & \tau>1\end{array}\right. \ .
  \label{eq: form factor COE}
 \end{eqnarray}

\begin{figure}[h]
  \centering
  \scalebox{0.8}{\includegraphics{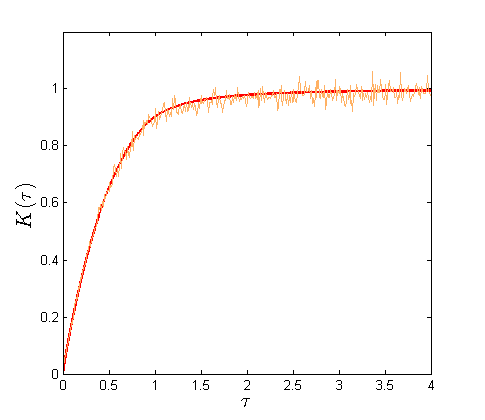}}
 \caption{The form factor $K(\tau)$ (unfolded spectrum) for $3$-regular graphs with 1000 vertices numerical \emph{vs.} the COE prediction.}
 \label{fig:Form Factors}
\end{figure}

The numerical data used to compute the nearest neighbor spacing
distribution $P(s)$, was used to calculate the form factor, as shown
in figure (\ref{fig:Form Factors}). The agreement between the
numerical results and the RMT prediction is apparent. This numerical
data triggered the research in \cite{IdanAmitUzy,IdanUzy2}.

The above comparisons between the predictions of RMT and the
spectral statistics of the eigenvalues of $d$-regular graphs was
based on the unfolding of the phases $\phi_j$ into the uniformly
distributed phases $\theta_j$. As will become clear in the next
section, it is more natural to study here the fluctuations in the
original spectrum and in particular the form factor
\begin{equation}
\widetilde{K}_{V}(t)=\frac{1}{V}\mathbb{E}\left(
\left|\sum_{j=1}^{V}{\rm e}^{it\phi_j}\ \right|^2\ \right)\ .
  \label{eq:raw form factor definition}
  \end{equation}
The transformation between the two spectra is effected by (\ref{eq:
definition theta_j}) which is one-to-one and its inverse is defined
by:
\begin{equation}
\phi = S(\theta) \doteq N_{KM}^{-1}\left (V\frac{\theta}{2\pi}\right
)\ .
 \label{eq:mapping}
\end{equation}
This relationship enables us to express  $\widetilde{K}_{V}(t)$ in
terms of $\ K_{V}(t)$. In particular, if $K_{V}(t)$ scales by
introducing $\tau = \frac{t}{V}$ then,
\begin{equation}
  \label{eq:K_tilde_and_K}
\hspace{-2 cm}\widetilde{K}(\tau=\frac{t}{V}) =
\frac{1}{\pi}\int_0^{\pi} d\phi K\left( \tau S^{'}(\phi) \right) =
2\int_0^{\pi/2} d\phi \rho_{KM}(\phi) K\left (\frac{\tau}{2\pi
\rho_{KM}(\phi)}\right).
\end{equation}
The derivation of this identity is straightforward, and is given in
\cite{IdanUzy2}.

With this summary of definitions and numerical data we prepared the
background for the main results of the present work, where we use
the trace formula to express the spectral form factor in terms of
the variance of the fluctuations in the counting of the number of
t-periodic nb walks. By assuming that the spectral fluctuations for
the graphs are given by RMT, we shall derive the variance-to-mean
ratio of t-periodic nb-walks on graphs. This approach is similar in
spirit to the work of Keating and Snaith \cite{KeatingSnaith} who
computed the mean moments of the Riemann $\zeta$ function on the
critical line, assuming that the fluctuations of the Riemann zeros
follow the predictions of RMT for the Circular Unitary Ensemble
(CUE).

\section{The Variance-to-Mean Ratio}
  \label{sec:var-to-mean}

\begin{figure}[h]
  \centering
  \scalebox{0.8}{\includegraphics{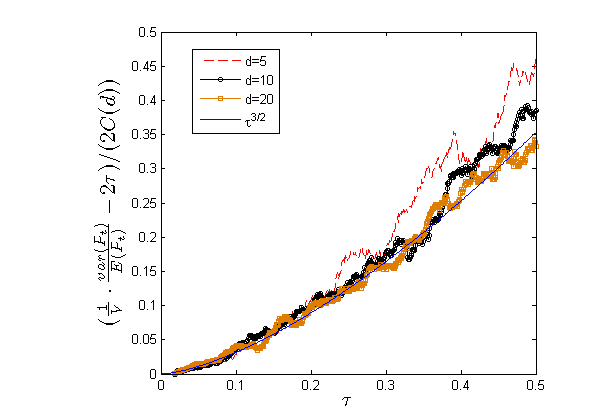}}
 \caption{$(\frac{1}{V}\cdot\frac{var(P_t)}{\mathbb{E}\left( P_{t} \right)
}-2\tau)/(2C(d))$ for various values of $d$ \textit{vs.} the curve
$\tau^{3/2}$.}
 \label{fig:Data Collapse}
\end{figure}

The spectral density (expressed in terms of the spectral parameter
$\phi$ (\ref{eq:specY})) is separated to its mean and fluctuating
parts:
\begin{equation}
\rho_R(\phi) = \rho_{KM}(\phi)+\tilde{\rho}(\phi)\ .
  \label{eq:spectral density phi}
\end{equation}
where $\rho_{KM}(\phi)$ is defined in (\ref{eq:Kesten McKay phi})
and:

\begin{equation}
\tilde{\rho}(\phi) = \frac{1}{\pi}\sum_{t=3}^\infty y_t \cos(t\phi)\
.
  \label{eq:fluctuating part phi}
\end{equation}
Using the orthogonality of the cosine, we can extract $y_t$,
\begin{equation}
y_t =  2\int_0^{\pi} \cos{(t \phi)} \tilde{\rho}(\phi) d\phi
\end{equation}
And so:
\begin{equation}
  \label{eq:(y_t)^2}
\mathbb{E}\left( y^{2}_t \right) =
 4\int_0^{\pi} \int_0^{\pi} \cos{(t \phi)} \cos{(t \psi)} \mathbb{E}\left( \tilde{\rho}(\phi)
\tilde{\rho}(\psi)\right) d\phi d\psi \ .
\end{equation}
From (\ref{eq:raw form factor definition}), we can write
$\widetilde{K}_{V}(t)$ equivalently as:
\begin{equation}
  \label{eq:K_tilde_normalized _to_unity}
\widetilde{K}_{V}(t) \equiv 2V \int_0^{\pi} \int_0^{\pi} \cos(t
\phi) \cos(t \psi) \mathbb{E}\left( \tilde{\rho}(\phi)
\tilde{\rho}(\psi) \right) d\phi d\psi\ ,
\end{equation}
and comparing (\ref{eq:(y_t)^2}) and (\ref{eq:K_tilde_normalized
_to_unity}) we get:
\begin{equation}
\widetilde{K}_{V}(t) = \frac{V}{2}\mathbb{E}\left( y^{2}_t \right)\
.
  \label{eq:K_tilde and y_t}
\end{equation}
Since asymptotically, $\mathbb{E}\left( P_t\right) = (d-1)^t$, and
using (\ref{eq:y_t}), equation (\ref{eq:K_tilde and y_t}) gives the
following remarkable equality between a spectral quantity and a
combinatorial one, for large $V$:
\begin{equation}
\widetilde{K}^{(A)}_{V}(t) =
\frac{1}{2V}\cdot\frac{\mathbb{E}\left(\left(P_{t}-\mathbb{E}\left(
P_{t} \right)\right)^2\right)}{\mathbb{E}\left( P_{t} \right)
}=\frac{1}{2V}\cdot\frac{var(P_t)}{\mathbb{E}\left( P_{t} \right) }
\ .
  \label{eq:K tilde and var over mean_P_t}
\end{equation}
This is the key ingredient in providing a closed formula for the
variance to mean ratio of nb-periodic walks.\\
The numerical evidence suggests that the form factor for graphs is
given by the COE expression (\ref{eq: form factor COE}). Then,
combining (\ref{eq:K tilde and var over mean_P_t}) and
(\ref{eq:K_tilde_and_K}), we get the desired formula for the
variance-to-mean ratio of periodic orbits:
\begin{equation}
  \label{eq:conjecture2}
\frac{1}{V}\cdot\frac{var(P_t)}{\mathbb{E}\left( P_{t} \right) } =
4\int_0^{\pi/2} d\phi \rho_{KM}(\phi) K_{COE}\left (\frac{\tau}{2\pi
\rho_{KM}(\phi)}\right)\equiv F_{COE}(\tau) \ .
  \label{eq:var_to_mean}
\end{equation}
The latter is the main result for this paper, providing a formula
for the variance-to-mean ratio for all values of $t$. The validity
of this result is supported by figure (\ref{fig:var-to-mean
numerics}) where the results of numrical simulations are displayed
together with the proposed function $F_{COE}(\tau)$. The simulations
were carried out by averaging over 100 random choices of graphs with
$V = 1000,\ d = 10$.\\
\begin{figure}[h]
  \centering
  \scalebox{0.8}{\includegraphics{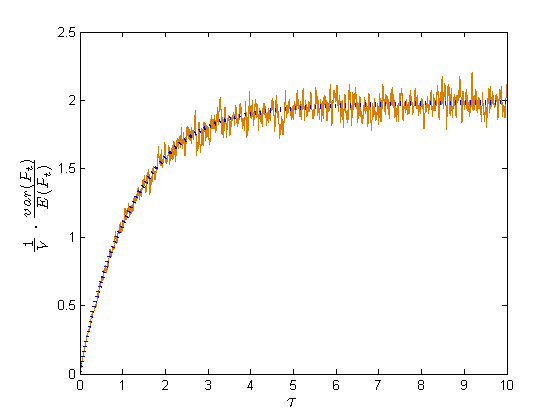}}
 \caption{$\frac{1}{V}\cdot\frac{var(P_t)}{\mathbb{E}\left( P_{t} \right)}$ accompanied by $F_{COE}(\tau)$ (dotted line).}
 \label{fig:var-to-mean numerics}
\end{figure}
The asymptotic behavior of (\ref{eq:conjecture2}) at the two
extremes of $\tau\rightarrow 0$ and
$\tau\rightarrow\infty$, can be obtained by using the known behavior of $K_{COE}$.\\
At $\tau\rightarrow\infty$, $K_{COE}\left (\frac{\tau}{2\pi
\rho_{KM}(\phi)}\right)=1$. Therefore we get that
\begin{equation}
\lim_{\tau\rightarrow\infty} F_{COE}(\tau)= 2
\end{equation}
This limit is apparent from figure (\ref{fig:var-to-mean numerics}).\\
At $\tau\rightarrow 0$, we can expand the middle part of
(\ref{eq:var_to_mean}) in powers of $\tau$ (see \cite{IdanUzy2} for
details). This expansion yields
\begin{equation}
  \label{eq:expansion in tau}
F_{COE}(\tau)=2\tau \cdot (1+C(d)\sqrt{\tau}+\ldots)
\end{equation}
where $C(d)$ is given explicitly by (\ref{eq:conjecture1}).\\
The most striking feature lies in the fact that the deviation from
the Poissonian expression is of order $\tau^{1/2}$. This is
illustrated in figure (\ref{fig:Data Collapse}) where
$(\frac{1}{V}\cdot\frac{var(P_t)}{\mathbb{E}\left( P_{t} \right)
}-2\tau)/(2C(d))$ is plotted for graphs with various values of $d$.
The expected power-law and data collapse are clearly visible for
$\tau<0.2$.\\

In our opinion, the numerical evidence presented above is convincing
enough to suggest that our main conjecture (\ref{eq:conjecture0}) is
valid. A combinatorial approach is called for to test it further.

\section*{Acknowledgments}

We thank Prof. Nati Linial for his continuous support and interest.
This work was funded by the Minerva Center for non-linear Physics,
the Einstein (Minerva) Center at the Weizmann Institute and the
Wales Institute of Mathematical and Computational Sciences) (WIMCS).
Grants from EPSRC (grant EP/G021287), and BSF (grant 2006065) are
acknowledged.

\end{document}